\documentstyle[12pt]{article}
\def\etal{{\it et. al.}}

\def\be{\begin{equation}}
\def\ee{\end{equation}}

\begin{document}
\baselineskip .3in
%\sloppy
\newpage
\pagestyle{plain}

\title{Frequency-dependent (ac) Conduction in Disordered Composites: a
Percolative Study}
%\author{Asok K. Sen$^*$ \\
\author{Asok K. Sen\thanks{asok@cmp.saha.ernet.in : Author for communication}\\
{\it LTP Division, Saha Institute of Nuclear Physics} \\
{\it 1/AF, Bidhannagar, Calcutta 700 064, India}\\
and \\
Abhijit Kar Gupta\thanks{abhi@imsc.ernet.in}\\
{\it Institute of Mathematical Sciences, CIT Campus} \\
{\it P.O. Taramani, Chennai 600 113, India}}

\date{\today}
\maketitle

\begin{abstract}
In a recent paper [Phys. Rev. B{\bf57}, 3375 (1998)], we examined in detail
the nonlinear (electrical) dc response of a random resistor cum tunneling
bond network ($RRTN$, introduced by us elsewhere to explain nonlinear response
of metal-insulator type mixtures).  In this work which is a sequel to that
paper, we consider the ac response of the $RRTN$-based correlated $RC$ ($CRC$)
model.  Numerical solutions of the Kirchoff's laws for the $CRC$ model give a
power-law exponent (= 0.7 near $p = p_c$) of the modulus of the complex ac
conductance at moderately low frequencies, in conformity with experiments on
various types of disordered systems.  But, at very low frequencies, it gives a
simple quadratic or linear dependence on the frequency depending upon whether
the system is percolating or not.  We do also discuss the effective medium
approximation ($EMA$) of our $CRC$ and the traditional random $RC$ network
model, and discuss their comparative successes and shortcomings.
\end{abstract}


%$^*${\it e-mail address: asok@cmp.saha.ernet.in}

\newpage

\section {\bf Introduction }

\vskip 0.1 in

Electrical response of a composite material made up of conducting and
insulating components to an external field is still of much interest.
To find the electrical conduction properties of random mixtures one usually
models a network with the notions of percolative processes \cite{sa, clerc}
as a basic framework. For simplicity one may consider a lattice-based
percolation and assume that the occupied bonds (with a probability $p$)
are metallic or good conductors and the rest of the unoccupied bonds (with
a probability $1-p$) are insulators or very poor conductors.  So it turns
out that one is dealing with a random electrical network where each bond
of the regular lattice constitutes an appropriate circuit element.  To 
study the linear direct current (dc) response of a random binary mixture,
one considers it to be a random (linear) resistor network ($RRN$) where
the conductance $g$ of a circuit element assumes the value $g_1$
with probability $p$, or $g_2$ (= 0 for insulators) with the complementary 
probability $1-p$. When one attempts to study the alternating current (ac) 
response of such a system at a frequency $f=\omega /(2\pi)$, the elementary
conductances $g_1$ and $g_2$ do, in general, become complex
quantities, termed as complex `admittances' (inverse of impedances),
because of the presence of inductances and/ or capacitances in some parts of
the circuit.  Electrical conductivities (both dc and ac) of various
composite systems, have been the subject of numerous studies \cite{clerc,
psp, bergstro, etopim4}.  On the theoretical side, analytical 
calculations involving related Maxwell's equations with appropriate boundary 
conditions, effective medium approximation ($EMA$), series expansion method,
Monte Carlo simulation, model random resistor network ($RRN$), etc.
are employed to study various interesting behaviors.

The key experimental facts to be contended with by a model in the case
of nonlinear dc response have been discussed in our recent works
\cite{skg, nldc}.  To reproduce the very low percolation threshold, a
`low'-temperature resistance minimum, and the nonlinear electric (dc)
response of metal-insulator composites, we had introduced in our original
work \cite{skg} a new semi-classical bond percolation model.  The
distinguishing feature in this model (compared to earlier $RRN$, etc.)
is the possibility of a local tunneling/ dielectric breakdown (beyond a
microscopic threshold voltage, $v_g$) between two metallic bonds
separated by a nearest neighbor lattice distance.  We call these specially
positioned insulating bonds, tunneling or `t-bonds'.  All the other insulating
bonds separated farther away remain insulating at any voltage however
large.  Clearly, since the positions of these t-bonds are
totally correlated to the positions of the metallic bonds in a particular
(random) configuration, this model may be considered as a {\it correlated
bond percolation} model.  The percolative phase transition aspects
\cite{cperc} and the dielectric breakdown aspects \cite{bdown} for this
model have already been discussed elsewhere.  Based upon this correlated
bond percolation, the circuit made of the random network of resistors and
t-bonds has been called a $RRTN$ (random resistor cum tunneling bond network).

In the present work which is a sequel to our work on the nonlinear dc
response of the $RRTN$ \cite{nldc} (referred to as I from now on), we are
concerned with the modeling of the generic ac response of a whole variety
of experimental systems within the $RRTN$ framework (reported briefly first
in the ref.\cite{skg}).  Experiments on the complex ac conductance
$G({\omega})$ in various composite systems, dispersed metals etc., as well as
many disordered/ amorphous systems \cite{clerc, psp,bergstro,etopim4,kkb,cbb,
hop}, report a non-integer power-law behavior of the modulus, $|G(\omega)|
= [(Re~G)^2 + (Im~G)^2]^{1/2}$.  At a fixed voltage and at a moderately
low-$\omega$, $[|G(\omega)| - G(0)] \propto \omega^{\alpha}$, where
$\alpha\cong$ 0.7 (the case of very low $\omega$'s is discussed later).
Now, at a low voltage $V$ (see I), $[G(V) - G(V=0)] \propto
(V - V_g)^{\delta}$ \cite {kkb}, where $V_g$ is the macroscopic threshold
voltage for dc nonlinearity
and the exponent $\delta$ is dimensionality-dependent.
There has been a recent claim \cite{kkb} for 3D carbon-wax composites
that $\delta = \alpha \cong $ 1.4.  We shall comment on this later.

In the traditional $RC$ network models used to study the ac response of
composites, the conducting bonds are pure (real) resistors and all the 
insulating bonds may behave as capacitors in the presence of an ac-field.
A fairly complete review and references (from the percolative aspect) on
the linear response may be found in the ref.~\cite{clerc}.  Another
more recent review in this regard is due to Shalaev \cite{shala}.

If one applies an ac electric field across our $RRTN$ model, a t-bond
between two nearby metallic bonds are expected to behave as a
capacitor.  Note that {\it perfect} capacitors at all the t-bonds
correspond to a situation where all the t-bonds have zero dc conductance
at low voltages,
and hence the $RRTN$ is in its {\it lower linear dc (or, ohmic) regime}.
Similarly, {\it leaky} capacitors with a very low constant conductance
for all the t-bonds, implies that the $RRTN$ is in its {\it upper linear
dc regime}.  To obtain the nonlinear ac response in the truly nonlinear
dc regime, one has to let the t-bonds be active or passive according to
the voltage differences across them.  In this work we study the
nonlinear ac response in either the upper or the lower linear dc regimes
only.  Further, we make a simplifying assumption that all the
capacitances across insulators farther than the nearest neighbor
distance are zero.  Based on the $RRTN$ model, this model for studying
ac response may thus be called a {\it correlated RC (CRC) model}.  We
believe that the simplicity of our model is physically appealing
and realistic enough.  Since the capacitors placed in this way never
percolates by themselves, the $|G(\omega)|$ is expected to connect
nonlinearly between its own lower
and upper ac saturation regimes of $\omega \rightarrow 0^+$ and
$\omega \rightarrow \infty$ respectively.

To make the above discussions more concrete we show in Fig.~1, a
typical numerical result for the real part of the ac conductance for
$p$ = 0.52 in a 20$\times$20 size sample with leaky capacitors.
There may be some genuine
concerns whether the upper saturation of the ac conductance does
actually occur in realistic samples.  In response we note that in most
of the early experiments, the steady frequency required to approach
the upper ac saturation regime were probably too high to be accessible.
Further, we find at least one experiment on $Li$-doped $NiO$ sample
\cite{pollak}, where the upper saturation is clearly observed.

As a pedagogical example (which will be useful for our interpretations
of the $EMA$ and the simulation results in the sequel), one may
calculate the complex $G(\omega)$ for some simple prototype circuits
using the capacitive conductance $g_t = j\omega c$ where $c$ is the
microscopic capacitance and $j = \sqrt{-1}$. For the case of Fig.~2(a):
\be
G(\omega) = {(r_1+r_2+\omega^2r_1r_2^2c^2) + j\omega r_2^2c \over
(r_1+r_2)^2 + \omega^2 r_1^2r_2^2c^2}.    \label{elem1a}
\ee
\noindent On the other hand, for the case of Fig.~2(b):
\be
G(\omega) = {\omega [\omega (r_1+r_2)c^2 + jc] \over
1 + \omega^2(r_1^2 + r_2^2)c^2}.    \label{elem1b}
\ee
\noindent It may be noted that the dc conductance $G(\omega = 0)$ of the
first circuit is $1/(r_1+r_2) > 0$ (thus it is a conductor), whereas
the dc conductance of the second circuit is zero (an insulator).  The
extremely low frequency ($\omega \rightarrow 0^+$) behaviors for both the
circuits is $[Re~G(\omega) - G(\omega = 0)] \propto \omega^2$ and $Im~
G(\omega) \propto \omega$.  Thus, $Re~G(\omega)$ or $Im~G(\omega)$ at
very low $\omega$ cannot distinguish between the two types of circuits.
But in the same limit, $[|G(\omega)| - G(\omega = 0)] \propto
\omega^2$ for the elementary {\it percolating} circuit of Fig.~2(a),
while $|G(\omega)| \propto \omega$ for the elementary {\it
non-percolating} circuit of Fig.~2(b).  As $\omega \rightarrow \infty$,
the upper ac saturation value for the Fig.~2(a) is $1/r_1$ and
that for the Fig.~2(b) it is $1/(r_1+r_2)$, both of which are finite
because the capacitor $c$ {\it does not geometrically extend} from one
electrode to the other.

Note that on adding together many such elementary circuits (i.e., on
increasing $L$), the rational algebraic function type behavior of
$Re~G(\omega)$ obtained from Eq.~(\ref{elem1a}) and Eq.~(\ref{elem1b})
changes over to a sigmoidal type function as shown in the Fig.~1 and
looks qualitatively very similar to the nonlinear dc conductance as a
function of $V$ \cite{skg, nldc}.  Here, $Im~G(\omega)$ is zero for
both $\omega = 0$ and $\infty$, with a broad peak in-between.  An
identical fitting function as the one for $G(V)$ in I, fits (solid
line) the simulation data (open circles) very well for six decades.
In the same spirit, $|G(\omega)|$ in the ac case is writen as:
\be
|G(\omega)| = G(\omega=0)+G_d(p)[1 - \exp(-\lambda\omega^{\mu})]^{\gamma}.
\label{gomega}
\ee
\noindent To obtain the power-law behavior, one linearizes the
exponential function in Eq.~(\ref{gomega}) for small $\omega$ (much
below the upper ac saturation regime) and one gets: $[G|(\omega)|
- G(0)] \propto \omega^\alpha$, where $\alpha = \mu\gamma$.
Now, for considering the ac response in our
$CRC$ model, we have virtually a three component mixture of
the ohmic bonds (conductance $g_o = g$), the t-bonds (in general leaky
with a complex $g_t = g_c + j\omega c$), and the insulating bonds
($g_i = 0$).  We assume that a sinusoidal voltage
$V = V_0\exp(j\omega t)$ is applied across the network.  We set $V_0 =
1$, $c = 1$ and $g_o = 1$ for convenience, thereby setting the scales
for the voltage, the frequency and the conductance respectively.  We shall
concentrate on either the $Re~G$ and $Im~G$, or the $|G|$ as a function
of $\omega$.  We would also study the phase-angle of the complex $G$
relative to the phase of the voltage source at any time $t$.

The rest of the paper is organized as follows.  In the Sec.~2, we
discuss the effective medium approximation ($EMA$) results for our $CRC$
model.  Next in the Sec.~3, we present the numerical Kirchhof's law
solution for the same model and compare them with the $EMA$ (for the
$CRC$) and some basic experimental results.  Finally, in the
Sec.~4, we summarize our findings with some concluding remarks.

\vskip 0.2 in

\section {\bf The EMA result for the CRC Model }

\vskip 0.1 in

For studying the ac conductance of the correlated $RC$ model within the
$EMA$, we assume for simplicity that the t-bonds behave as {\it perfect}
capacitors ($g_t = j\omega c$).  So, we are in the lower linear dc regime.
The probabilities of a bond to be ohmic, tunneling (capacitive) or purely
insulating in a 2D square lattice are:
\begin{eqnarray}
P_o & = & p, \\                                      \label{ema1} 
P_t & = & (p^3+3p^2q+3pq^2)^2q,\\                    \label{ema2}
P_i & = & 1 - P_o - P_t = [1-(p^3+3p^2q+3pq^2)^2]q,  \label{ema3}
\end{eqnarray}
where $q=1-p$.  The $EMA$ equation for the effective (i.e., {\it
average}) complex conductance $G_e$ for this three component system
may be written as \cite{nldc, kirk}:
\be
AG_e^2 + BG_e + C = 0,  \label{quad}
\ee
where $A = (d-1)^2$, $B = (d-1)[(1-dP_o)g_0 + (1-dP_t)g_t]$ and $C = -[(d-1) -
dP_i]g_og_t$~.  The solution of the quadratic Eq.~(\ref{quad}) is
\be
G_e = {- B \pm (B^2 - 4AC)^{1/2} \over 2A}.  \label{effcon}
\ee
We show below that only the `+' sign in front of the square-root is the
physical one for our purpose and henceforth
use that only.  In passing we note that the new percolation threshold
for our $RRTN$ on a square lattice is $p_{ct}$ = 1/4 using the $EMA$
(see I) and $p_{ct} \cong 0.181$ \cite{cperc} using a real space
renormalization \cite{gene} with finite size scaling analysis. 

Separating the real and the imaginary parts of $G_e(\omega)$, one gets
\be
Re~G_e(\omega) = {(2P_o-1)\over 2} + {1 \over {2\sqrt{2}}}
[X+(X^2+Y^2)^{1/2}]^{1/2} ,                              \label{recon}
\ee
and,
\be
Im~G_e(\omega) = {\omega(2P_t-1)\over 2} + {1\over {2\sqrt{2}}}
[-X+(X^2+Y^2)^{1/2}]^{1/2},                                 \label{imcon}
\ee
where $X = (2P_o - 1)^2 - \omega^2(2P_t - 1)^2$ and $Y = 2\omega[(2P_o 
- 1)(2P_t - 1) - 2(2P_i - 1)]$.  It may be noted here that $P_t(p)$
has a single broad peak structure with a maximum
value of about 0.3840 at a $p$ = 0.4800.  Thus, the quantity 
$\omega(2P_t - 1)$ is always negative.  Further, $X$ is also negative
for $\omega > |(2P_o-1)/(2P_t-1)|$ and approaches $-\infty$
quadratically as $\omega \rightarrow \infty$.  We will take only the
absolute value of the square-rooted expression for two reasons; (i)
$Re~G_e(\omega)$ in Fig.~3 achieves the necessary upper ac saturation
since the built-in square-root function in the computer uses exactly
that, and (ii) this procedure keeps $Im~G_e(\omega) > 0$ for all
$\omega > 0$ (needed since $Im~g_t > 0$). 
 
Let us first check the $\omega \rightarrow \infty$ limit.  The
dissipative part of the complex conductance, $Re~G_e(\omega)$, should
be positive, finite, and greater than $G_e(\omega = 0)$ in this limit.
Further, the reactive part $Im~G_e(\omega)$ of the $CRC$
network must become zero (i.e., show no response) when the driving
field oscillates much faster than the network's relaxation time (or,
the {\it time-constant}).  Now, in this limit,
\be
Re~G_e(\omega) = {(2P_o-1)\over 2} + {1 \over 2}~
|{2(2P_i-1) \over (2P_t-1)} - (2P_o-1)| ,       \label{ginfr}
\ee
and,
\be
Im~G_e(\omega) = {\omega(2P_t-1)\over 2} + {1\over 2}~|\omega (2P_t-1)|,
                                                        \label{ginfi}
\ee
and thus for all $p > p_{ct}$, $G_e(0) < Re~G_e(\omega=\infty) <
\infty$.  Also, clearly $Im~G_e(\omega)$ in this limit is zero.  Thus,
both the conditions hold in the $\omega \rightarrow \infty$ limit.

Next we check the asymptotic expansion as $\omega \rightarrow 0^+$.  In
this limit, $[Re~G_e(\omega) - G_e(0)] \sim \omega^2$ and $Im~G_e(\omega)
\sim \omega$.  We find that this extremely low-$\omega$ behavior of the
complex $G_e$ in the $EMA$ is generic for all $p_{ct} < p < 1$ ({\it except}
at $p=p_c$, see below), and $[|G_e(\omega)|-G_e(\omega=0)] \sim \omega^2$.
Hence the $EMA$ cannot distinguish between a percolating and a
non-percolating configuration from their low-$\omega$ behavior.

Now, in the special case when $p = p_c$ (= 1/2 for a square lattice), we
find that in the limit $\omega \rightarrow 0^+$, both the $Re~G_e(\omega)$
and $Im~G_e(\omega)$ varies as $\omega^{\alpha}$, where $\alpha$ = 0.5.
Obviously,
$|G_e(\omega)| \sim \omega^{0.5}$ in this limit.  In passing we would like
to quote the extremely low-frequency $EMA$ exponent in the
case of 3D.  By looking at the $EMA$ expressions for $G_e(\omega)$ which are
the analogues of Eqs.~(\ref{recon}) and (\ref{imcon}) for a simple cubic
lattice at its $EMA$ percolation threshold ($p_c$ = 1/3 in 3D), one finds
again that $\alpha = 0.5$ (in 3D).

We show in Fig.~3 a log-log plot of $Re~G_e(\omega)$ against $\omega$ in
2D [Eq.~(\ref{recon})].  In conformity with the asymptotic expansions
obtained above analytically for very small $\omega$'s, the $EMA$ results 
shown in the Fig.~3 give $\alpha(p) = 2.0$ for all $p_{ct} < p < 1.0$
except for the special case of $p=p_c$.  We have $\alpha(p_c) = 0.5$
both in the very low and in the moderately low-$\omega$ regimes much below
the upper saturation of $Re~G_e(\omega = \infty)$.  Further for each fixed
$p\ne p_c$, there is a characteristic $\omega_0(p)$ around which $\alpha(p)$
starts crossing over from 2.0 to the moderately low-$\omega$ exponent of
about 0.5.   The jump of $\alpha(p)$ at an extremely low-$\omega$ from
2.0 to 0.5 is the hallmark of the inadequacy of $EMA$.  One does also
note that for a fixed $p \ne p_c$ , this crossover region becomes
smaller for smaller $p$'s.  Indeed, the crossover region finally tends
to vanish as $p \rightarrow p_{ct}$ ($\simeq$  0.18 in 2D).  Thus, we
observe that there are no low to moderate $\omega$ crossovers for
for $p = p_c$ with $\alpha = 0.5$ and similarly for $p = p_{ct}$
where $\alpha = 2.0$. 

We did also calculate the $EMA$ results
with leaky capacitors at each t-bond ($g_t=g_c + j\omega c$), i.e., in
the upper linear dc regime.  The only change here compared to the case
of perfect capacitors above is that even at $p=p_c$, $|G_e(\omega)| =
G_e(\omega=0) + k\omega^2$ ($k$=constant) for very low-$\omega$.  Also
for all $p>p_{ct}$, $G_e(\omega=0) > 0$, as expected.

We would also like to see the behavior of phase-angle of the complex
conductance $G_e(\omega)$ with respect to the frequency ($\omega$). 
The phase-angle ($\phi$) is defined through
\be
\tan{\phi} = {Im~G_e(\omega) \over Re~G_e(\omega)}. 
\ee
In the Fig.~4, we plot the phase-angle ($\phi$) of the complex conductance
against frequency ($\omega$) for the $EMA$.  As expected, this angle is
zero [just like the $Im~G_e(\omega)$] both at very small and at very large
$\omega$'s.  Further the phase has a
peak value $\phi = \phi_m$ which increases as $p$ is decreased and the
$\omega$ at which the peak occurs is $p$-dependent.  We find from 
this Fig.~4 that the positions of the peaks tend towards zero and that
$\phi_m$ becomes progressively larger as $p$ approaches $p_c$ from higher
values.  It may be noted that we cannot calculate the phase angle $\phi$
for $p < p_c$ in the $EMA$ with confidence because the quantity
$G_e(\omega = 0) = 2P_o - 1$ takes on {\it unphysical negative} values.
Hence we have not shown any curve for $p < p_c$ in Fig.~4.

\bigskip

\section {\bf The Numerical Result for the CRC Model}

\smallskip

We now solve Kirchoff's laws in our 2D complex network at each node of our
correlated $RC$ model.
We obtain the complex conductance of the macroscopic samples, their
real and imaginary parts, the $modulus$ values and the phase-angle through
iterative numerical solution using the Gauss-Seidel relaxation method. 
In the Fig.~5, we have plotted the modulus of the average complex 
conductance, $|<G(\omega)>|$ against $\omega$ for $0.3 \le p \le 0.7$ and for
an external sinusoidal voltage $V = cos(\omega t)$.  We let the $t$-bonds be
{\it leaky} capacitors with a conductance $g_c = 0.001 << g_o$ and averaged
over the same twenty configurations for each $\omega$.  So, we are virtually
in the upper dc saturation regime.  For clarity, we have shown the graphs in
Fig.~5 from zero to a moderately low $\omega$ (=0.1) much below the upper ac
saturation regime.  We find that for
$p \simeq p_c$~, the respective graphs in Fig.~1 and Fig.~5 are fitted very
well with the Eq.~(\ref{gomega}) such that $\alpha = \mu\gamma \simeq 0.7$.
Thus, in the moderately low-$\omega$ regime, $\alpha$ obtained
experimentally in varieties of systems \cite{cbb, long}, and
in an `Extended Pair Approximation' (EPA) theory \cite{sumbut}
is close to what we obtain here (i.e., 0.7).  We remark here
that for a 2D system exactly at $p_c$~, $\alpha = 0.5$ both in the simple
$RC$ model (see e.g., Ref.\cite{clerc}) and the $EMA$ of our
correlated $RC$ model.  Indeed, both of them fall short of the realistic
value of $\alpha$ not only at $p = p_c$, but also at any $p \ne p_c$.
Further, these theoretical and experimental results do not support any
evidence for the statement that $\alpha = \delta$ \cite{kkb}.  For example,
for our $RRTN$ model near $p_c$ in 2D, $\delta \simeq 1$, whereas $\alpha
\simeq 0.7$.  Indeed, as we had commented recently \cite{comment}, there is
no good physical argument supporting the equality.  The main point in the
comment is that since the actual experimental data become flat at large
$\omega$, it is quite well-known that the exponent obtained may depend
crucially on the arbitrarily chosen (i.e., without any particular physical
reason) upper cut-off in $\omega$ for fitting purposes.  Consequently,
the possibility of an unique function should be explored (as we do here)
which fits all the way from the lower to the upper saturation
range (if sufficient data are available), and then the power-law behavior
in the moderately low-$\omega$ range should be extracted from that function. 

Now, as we had mentioned before, we know of at least one experiment by Pollak
\etal \cite{pollak} on $Li$-doped $NiO$ single crystals at sufficiently low
temperatures, where the upper saturation of the $Re~G_e(\omega)$
may be clearly observed.  The frequency range used in that work is from about
$10^7~Hz$ to about $10^{10}~Hz$.  We did not try to fit them by our method
since no data below $10^7~Hz$ was available.  In any case we note that the
upper saturation is also consistent with the fact that the measured
relaxation time for this sample is about $2.2\times 10^{-10}$ s.  In passing
we do also note that in this system as well, the doping concentration of $Li$
is extremely low: from about $13\times 10^{-6}$ to about $136\times 10^{-6}$,
and that the pristine $NiO$ is an insulator.
We observe that the upper saturation value of the conductivity for this
experiment \cite{pollak} is about 0.1 S/m, and that the data for
all the three concentrations used for $Li$, fall closely enough to their
fitting function in the high frequency range, but not so well in the low/
moderate frequency range.  The power-law exponent $\alpha$ (around $10^7 Hz$)
seems to be close to 2.0.

Next, we we would like to emphasize that the apparently excellent fitting
shown in Fig.~1 with Eq.~(\ref{gomega}) may be misleading to the eye
in the very low-$\omega$ regime.
A careful analysis of data in a more revealing log-log plot shows that for
very low $\omega$, the fittings may be actually quite bad.  We observed for
the two elementary circuits (in Fig.~2) for very low $\omega$, $|G(\omega)|
\sim \omega^2$ or $\omega$ depending upon whether the circuit is
percolating or not.  Similarly, we anticipate that for extremely low
$\omega$ and in the lower dc regime, the $CRC$ network would also show such
simple behaviors (instead of a more complicated percolative behavior
in the moderately low-$\omega$ regime as described above).  Further this
simple behavior is expected to persist (as seen in our $EMA$ results)
generically for each $p$ up to some scaled crossover frequency $\omega_0$
depending on $p$, $g_o$, $g_c$ and $c$.  For $\omega > \omega_0$, we expect
macroscopic percolative effects to gain control and instead of
Eq.~(\ref{gomega}), $|G(\omega)|$ should follow a more general equation
closer in form to that used for the dc-conductance [cf. Eq.~(4.7) of I]:
\be
|G(\omega)| = G(\omega_0) + G_d(p)[1 -
\exp(-\lambda[\omega-\omega_0]^{\mu})]^{\gamma}.
\label{acfit}
\ee
A typical fit by the above equation is shown in Fig.~6 for a system size
$L = 20$ and  $p$ = 0.3  for about seven decades in $\omega$ using leaky
capacitors.  In the inset of Fig.~6, we show the log-log plot for the
$modulus$ value of $G$ and observe a very convincing quadratic behavior
with $|<G(\omega)>| = 0.01219 + 55\omega^2$ for $\omega_0
\le 0.01$.  This observation is totally matching with our analyses for
Fig.~2(a).  For $\omega > \omega_0$, we get a very
good fit for the intermediate frequency range with the general
Eq.~(\ref{acfit}) where $\alpha = \mu\gamma \simeq 3.0$ (which is much
larger than 0.7).

Next in the Fig.~7, we show another fit for a single configuration
(at $p=0.45 < p_c$) with perfect capacitors at the t-bonds ($g_c = 0$,
non-percolating) of the $CRC$ network with $L$=20.  In this case one can
easily observe that in the very low-$\omega$ range up to a crossover
frequency $\omega_0 \simeq$ 0.01, $|G(\omega)| \simeq 6\omega$.  This
behavior is akin to that of the elementary circuit of Fig.~2(b).  Further,
beyond $\omega_0$~, we do again have an excellent fitting with the
Eq.~(\ref{acfit}) with $\alpha = \mu\gamma \simeq 0.5$.  So we have two
things to note here for non-percolating configurations (with perfect
capacitors, i.e., in the lower dc regime):
(i) the very low-$\omega$ behavior is linear in $\omega$,
and (ii) the intermediate frequency behavior seems to give a lower value
for the exponent compared to that for percolating configurations (e.g., as in
the upper dc regimes of Fig.~1 and Fig.~6 where $\delta =$ 0.7 near $p_c$).
Note that for the non-percolating configurations (only at $p_c$) in the lower
linear dc regime, $\alpha = 0.5$ in the case of $EMA$ as well.

For many practical situations, this intermediate frequency range
(away from both the lower and the upper ac saturation regimes) is
of prime interest.  For t-bonds with leaky capacitors at any
$p > p_{ct}$ (i.e., in the upper dc regime), we find that $\alpha$ has a
minimum value of
about 0.7 near $p_c$, and increases on both sides of it with a value
of about 3.0 at $p$ = 0.3 (as shown above) and of about 1.35 (not explicitly
shown here) at $p$ = 0.7.  Clearly, this result has a qualitative similarity
with the dc nonlinearity exponent $\delta(p)$ as a function of $p$ as shown
in the Fig.~11 of I.

Finally, the variation of the phase-angle ($\phi$) with frequency ($\omega$)
obtained by numerical (Kirchoff's laws) solution of our $CRC$ network
has been shown in the Fig.~8 for 0.3 $\le p \le $ 0.7.
One can easily observe from this figure that in the $CRC$ model, for
configurations with $p$ around $p_c$ for a 2D square lattice, the peak
value of the phase, $\phi_m \cong 0.7$ (radian).  This is close to the
{\it universal phase-angle value} of $\pi/4$ radian obtained in the simple
$RC$ model in 2D at $p_c$ as predicted by Clerc \etal \cite{clerc}.  As
noted before, the $EMA$ calculations (i.e., mean-field approach) of the
phase $\phi(\omega)$ for our $CRC$ model also obtains $\phi_m \cong 0.7$ rad
when $p \cong p_c$.  The main problem with the $EMA$ in this case is that
the peak occurs at an $\omega$ of about 0.1 rad/s (instead of at around
0.2 rad/s as for the $CRC$ in Fig.~8 above) and that it is too broad.
Further, as shown in the Fig.~8 for the $CRC$ network which is expected
to be much more realistic (better than the mean-field theory, for one
reason) from the demonstrations and the arguments above, $\phi_m$ increases
as $p$ decreases: from a value of $\phi_m \simeq 1.1$ rad for $p$ = 0.3 to
a value of $\phi_m \simeq 0.4$ rad for $p$ = 0.7.

\vskip 0.2 in

\section {\bf Summary and Conclusion}

\vskip 0.1 in
In this work we have studied the nonlinear ac response (in the two
linear dc regimes) of a variety of materials, e.g., a binary
metal-insulator type composite, a dispersed metal, amorphous
semiconductors at high field, etc. by modeling them in terms of our
recently proposed $RRTN$ \cite{skg, nldc} model.  This $RRTN$ (Random
Resistor cum Tunneling-bond Network) driven by an ac electric field
gives rise to a correlated $RC$ ($CRC$) model because we place the
capacitors only at the t-bonds whose positions are correlated to the
positions of the nearest neighbor gaps between two ohmic bonds, as
discussed above.  For simplicity, the rest of the insulating bonds are
assumed to remain inactive at any voltage however large.  Thus this
model effectively behaves like a three component mixture.  We studied
the complex ac conductance $G(\omega)$ of our $CRC$ model as a function
of $\omega$ and at a fixed low voltage both by using an effective medium
approximation ($EMA$) and by an iterative numerical solution using
Kirchoff's laws.  The real part $Re~G(\omega)$ or the modulus
$|G(\omega)|$ behaves qualitatively very similarly to that for the real,
nonlinear dc conductance $G(V)$ as a function of the external voltage
(see I).  Thus we fit the $Re~G(\omega)$ or the $|G(\omega)|$ with an
exponential-related function as in Eq.(\ref{acfit}).  The fitting with
this function seems to be very good just like the fitting of $G(V)$ with
a similar function is optimum.  But the similarity ends there.

Whereas there is only one exponent for the power-law type low voltage dc
response, there are two different power-law regimes for, say $|G(\omega)|$,
much below its ac upper saturation regime.  Since any realistic sample
should not be able to respond to an extremely large $\omega$,
$Im~G(\omega)$ must vanish as $\omega \rightarrow \infty$, and that does
automatically happen in our $CRC$ model.  Thus $|G(\omega=\infty)|$ is real
and finite.  For non-percolating systems at very low-$\omega$, $|G(\omega)|
\sim \omega$, whereas if percolating, $[|G(\omega)| - G(0)] \sim \omega^2$.
This happens irrespective of the value of $p$.  Next for $\omega >
\omega_0$, i.e., at moderately low $\omega$'s (still much below the upper
ac saturation regime), the system crosses over to a non-integer power law
behavior with a positive exponent $\alpha(p)$ which depends on the value of
$p$.  If we use leaky capacitors at the t-bonds, the system would percolate
for all $p \ge p_{ct}$ ($\simeq$ 0.18) \cite{nldc}.  For such situations,
if $p \simeq p_c$, $\alpha$ has the minimum value of about 0.7 and this
value matches with that for many experiments.  Away from $p_c$ (both for
higher and lower values), with leaky capacitors at the t-bonds, $\alpha(p)$
becomes progressively larger than 0.7.  Now, if we use perfect capacitors,
then for non-percolating configurations close to $p_c$, $\alpha(p) \simeq
0.5$ which is significantly lower than 0.7.  It may be noted that both
the $EMA$ and the traditional $RC$ model cannot obtain anything other than
the value of 0.5 for $p$ close to $p_c$.  Further, in the case of $EMA$,
the exponent $\alpha(p)$ = 2.0 for any $p \ne p_c$.

We did also look at the phase angle $\phi(\omega)$ of the complex $G$ at
several $p$'s.  Since $Im~G(\omega)$ vanishes both at very high and very
low $\omega$'s (at any $p$), $\phi(\omega)$ would also have this
property as a function of $\omega$ with at least one peak value ($\phi_m$)
somewhere in-between.  There is one curve for each $p$ with a different
peak value.  In the case of $EMA$, $\phi(\omega)$ cannot be calculated
for any $p < p_c$ and so in Fig.~4, we have shown them only for $p > p_c$.
For the $CRC$ network, we do not have this pathology, and hence in the
Fig.~8, we have shown it for 0.3 $\le p \le$ 0.7.  In general, $\phi_m$
increases both in the $EMA$ and our numerical solution of the $CRC$
model as $p$ decreases.
For the latter, it is about 0.4 rad at $p = 0.7$ and 1.1 rad for $p = 0.3$.
It is interesting to note that at $p = p_c$, $\phi_m$ takes the universal
value of about $\pi/4$ rad in the simple $RC$ model, numerical solution
and the $EMA$ of our $CRC$ model.   But, whereas the $\phi_m(p_c)$ occurs
at an $\omega = 0.2$ rad/s for the $CRC$ model, it occurs at an $\omega =
0.1$ rad/s in the case of the $EMA$.   Further the peak around $\phi_m$
is too much broad for the $EMA$.

\vskip 0.2 in

\noindent {\bf Acknowledgement }

\vskip 0.1 in
We thank A. Hansen, H. Herrmann, S. Roux, B.K. Chakrabarti, K.K.
Bardhan, R.K. Chakrabarti and U.N. Nandi for many useful discussions
(the last three on their experiments) at the early stages of this work.
AKS acknowledges the support of the A.S. ICTP, Trieste, Italy, where
the manuscript was thoroughly re-written.

\newpage

\noindent {\bf Figure Captions:}

\bigskip

{\bf Fig.~1} Real part of the complex $G(\omega)$ for our model correlated
$RC$ ($CRC$) network at $p = 0.52$, with leaky capacitors placed at the tunnel
junctions. The solid line given by Eq.~(\ref{gomega}) appears to give a very
good fit.

\bigskip

{\bf Fig.~2}  Two prototypical elementary circuits each with two ohmic resistors
and one capacitor.  Note that the Fig.~2(a) corresponds to a metal-like
zero-frequency behavior (i.e., $G_e(\omega = 0) > 0$), whereas the elementary
circuit in Fig.~2(b) corresponds to an insulator (i.e., $G_e(\omega = 0)$ = 0).

\bigskip

{\bf Fig.~3} The $EMA$ result in 2D for the $CRC$ model: real part of the effective
conductance $Re~G_e(\omega)$ against $\omega$ for a set of values of $p$.

\bigskip

{\bf Fig.~4} The $EMA$ result for $CRC$ model in 2D: the phase-angle ($\phi$)
against $\omega$ for different values of $p$.

\bigskip

{\bf Fig~.5} The numerical results in 2D for the $CRC$ model with leaky capacitors
with a finite real conductance $g_c > 0$: the $|<G(\omega)>|$ against $\omega$ for
a set of values of $p$.  For each $p$ and $\omega$, the average over 20 different
configurations were taken. 

\bigskip

{\bf Fig.~6} The $|<G(\omega)>|$ for $p$ = 0.3 and a square lattice of size
$L$ = 20.  Just as in Fig.~5, each t-bond represented a leaky capacitor and
an average over 20
configurations were taken. The inset shows an $\omega ^2$-dependence upto a
crossover frequency of $\omega_0 \simeq 0.01$.  Above $\omega_0$, the
Eq.~(\ref{acfit}) gives a very good fit as shown.

\bigskip

{\bf Fig.~7} Another example of $|G(\omega)|$ against $\omega$ for $p$ = 0.45
and for a typical configuration on a square lattice of size $L$ = 20. The very
low-$\omega$ part shows a purely linear behavior. But for $\omega > \omega_0
\simeq 0.01$, Eq.~(\ref{acfit}) gives the optimum fit.

\bigskip

{\bf Fig.~8} Phase-angle ($\phi$) as a function of $\omega$ from the numerical
solutions of our $CRC$ model. 

\end{document}